\documentclass[twocolumn,aps,amsmath,nofootinbib,english]{revtex4-1}
\usepackage{graphicx}
\usepackage{amssymb}
\usepackage{xspace}
\usepackage{amsfonts}
\usepackage{color}

\def\beq{\begin{equation}}
\def\eeq{\end{equation}}
\def\beqn{\begin{eqnarray}}
\def\eeqn{\end{eqnarray}}
\makeatother
\begin{document}

\title{Deformed Special Relativity from Asymptotically Safe Gravity}
\bigskip
\author{Xavier Calmet}
\email{x.calmet@sussex.ac.uk}
\affiliation{Physics and Astronomy, 
University of Sussex,   Falmer, Brighton, BN1 9QH, UK}
\author{Sabine Hossenfelder}
\email{hossi@nordita.org}
\affiliation{NORDITA, Roslagstullsbacken 23, 106 91 Stockholm, Sweden}
\author{Roberto Percacci}
\email{percacci@sissa.it}
\affiliation{SISSA, via Bonomea 265, I-34136 Trieste
and 
INFN, Sezione di Trieste, Italy}
\medskip
\begin{abstract}
By studying the notion of a fundamentally minimal length scale in asymptotically safe gravity we
find that a specific version of deformed special relativity ({\sc DSR}) naturally arises in this approach. We then consider two thought
experiments to examine the interpretation of the scenario and discuss
similarities and differences to other approaches to {\sc DSR}. 
\end{abstract} 

\maketitle

\section{Introduction}

A modification of the dispersion relation is believed to be one of the most likely low-energy signatures
of quantum gravity and the prospect of detecting such effects in
the near future has generated wide interest. Deformed Special Relativity, also known as Doubly 
Special Relativity or {\sc DSR} \cite{AmelinoCamelia:2000ge},
was originally conceived as a way of modifying the dispersion relations
of relativistic particles without breaking Lorentz invariance, i.e.\ without introducing a preferred
frame. This modification is motivated by wanting the Planck energy to be an invariant under
a change of reference frame, which cannot be achieved with the usual Lorentz-transformations. 
{\sc DSR} is commonly interpreted as a phenomenological description
of presently unknown quantum gravity effects. A derivation of {\sc DSR} from a 
proposed theory of quantum gravity is not yet available, at least not in $3+1$ dimensions.

The core idea of {\sc DSR} is that the Lorentz group
is realized nonlinearly on the four--momenta in such a way as to preserve the Planck mass, $m_{\rm p}$, as 
a maximal energy scale.
 This can be achieved by noting that the standard Lorentz-transformations,
$\Lambda$, acting on the momentum four--vector, ${\bf p} = (p^0, {\vec p})$, does have two invariants: zero and infinity.
To arrive at a modified transformation that leaves invariant a finite energy, one introduces
a function ${\bf k}=F({\bf p})$ with the property that $F^{\mu}(\infty) = m_{\rm p}$, for either the
energy component or the three--momentum or both. Since ${\bf p}$ transforms under
the usual Lorentz-transformation into ${\bf p}' = \Lambda {\bf p}$, one finds the transformation for ${\bf k}$ 
\beqn
{\bf k}' = F(\Lambda {\bf p}') = F (\Lambda F^{-1}({\bf k})) \quad,
\eeqn
which then has the desired property of leaving invariant $m_{\rm p}$. The resulting modified 
Lorentz-transformations can be explicitly constructed this way. They
have the feature that they keep invariant two scales now: the Planck mass and a constant $c$
of dimension velocity. The concrete realization of {\sc DSR} then depends on the choice of the function $F$, as
well as on the interpretation of the quantities ${\bf k}$ and ${\bf p}$. 

The general properties of $F$ are that for small values of ${\bf p}$ it should reduce
to the identity, while at large values it should approach a finite constant which
was the original motivation. The function should moreover be invertible. In the literature, 
different concrete realizations for the function $F$ have been proposed, leading
to different models. The rotation subgroup is usually preserved in its linear form,
and $F$ is assumed to depend on the three-momentum only via
its modulus $|\vec p|$,
and the three components $F^i$ are assumed to be the same. 
In many cases it is assumed that $F$ does not depend at
all on three momentum, but only on energy $E=p^0$. The modified
dispersion relation follows from the choice of $F$. These types
of models can, but need not necessarily result in an energy-dependent speed
of light (more precisely: the speed of massless particles). They do imply
a generalization of the uncertainty principle as a manifestation of the
presence of a maximal energy scale. The relations between
modified dispersion relations, deformed Lorentz-transformations, a 
generalized uncertainty principle, and an energy-dependence of the speed
of light have been laid out in \cite{Hossenfelder:2005ed}.
In case the speed of light is energy-dependent, the invariant constant
$c$ of the deformed Lorentz-transformations is the speed in the low-energy limit.

In the more common {\sc DSR}
interpretation \cite{AmelinoCamelia:2000ge} ${\bf k}$ is considered the true momentum four--vector, whereas the
normally transforming ${\bf p}$ is a ``pseudo--momentum'' without physical meaning. 
In the interpretation of \cite{Hossenfelder:2003jz} instead, ${\bf p}$ is still the physical four--momentum while
${\bf k}$ is the wave-vector that is now related to the momentum non-linearly, corresponding
to an energy-dependence of Planck's constant. These
interpretations have been compared in \cite{Hossenfelder:2006cw}. 

The standard interpretation of {\sc DSR} has proven to be difficult to
extend to a more complete theory; if dealing
with a version that has an energy-dependent speed of light difficulties increase. 
The non-linear transformation of four--momenta necessarily leads to an unusual addition
law for energies which in turn leads to problems with constructing
multi-particle bound states, known as the ``soccer-ball problem'' \cite{Judes:2002bw,Liberati:2004ju}.
An energy-dependent yet observer-independent speed of light causes
problems with locality \cite{Schutzhold:2003yp}. 
Presently, for neither of these problems a satisfactory solution is known. 

In this brief paper, we want to look at the interpretation of {\sc DSR} from
a new point of view. We will see that {\sc DSR} 
occurs naturally in scenarios with asymptotically safe gravity, and that
this provides us with an
unambiguous interpretation. 

 In the
following we will use units with $\hbar = c = 1$ and $G=1/m_{\rm p}^2$.

\section{The Running Planck Mass}

At the beginning, let us revisit the original motivation for {\sc DSR}, to render
 the Planck mass invariant under Lorentz-transformations. 

The Planck mass determines the strength of the 
gravitational interaction $G=1/m_{\rm p}^2$. As a coupling constant, $G$ is however expected to run 
with energy as do all other coupling constants, thus $m_{\rm p}$ becomes energy-dependent too. 
One is thus left to wonder what is the meaning of the invariant constant $m_{\rm p}$ in the deformed
Lorentz-transformations. It cannot correspond to the physical coupling
constant at all energies. It can only be interpreted as being related to the coupling constant in the low
energy limit, much like the other constant in the transformation, $c$,
corresponds to the speed of light in the low energy limit while the
physical speed of light actually becomes energy-dependent. 

But more
importantly, since the running Planck mass is the physically relevant
one, we have to ask now for the meaning of the quantities we are
transforming. Introducing an algebra that preserves the Planck mass seems meaningless if the Planck mass runs with energy. Moreover, as any other renormalized quantity, its value at a given energy is dependent on the renormalization scheme which is used. The situation is similar to Quantum Chromodynamics ({\sc QCD}) where the value of $\alpha_s$ at some energy scale is scheme dependent at higher order. In {\sc QCD} the physical parameter is the {\sc QCD} scale, which defines the energy scale at which QCD becomes strongly coupled. The analog of the {\sc QCD} scale for quantum gravity is $\mu_\star$ defined by  $G(\mu_\star) \sim \mu_\star^{-2}$ \cite{Calmet:2008tn}. This condition implies that fluctuations in spacetime geometry at length scales $\mu_*^{-1}$ will be unsuppressed.  In a consistent theory of quantum gravity, $\mu_\star$, i.e.\ the scale at which quantum gravity effects become relevant,  might be the physical quantity which one might want to preserve under deformed Lorentz transformations.

The way we will work towards understanding the implications of the running
coupling is to emphasize that physical observables should be dimensionless, and
these dimensionless constants have to be constructed using physical
units, now given by the energy-dependent Planck mass. This basically
means that at high energies, when the energy-dependence of the Planck
mass becomes relevant, we have to be careful to chose the physically
meaningful quantities. These considerations will allow us to address the 
above mentioned interpretational
questions in {\sc DSR}. 

To be more specific, we will here deal with a particular scheme for the
running of the gravitational coupling that has recently attracted a lot
of attention, namely asymptotically safe gravity. Before we come to our
core argument, in the next section we will briefly summarize the main
features of asymptotically safe gravity.

\section{Asymptotically Safe Gravity}

Asymptotic Safety is generally understood as a possible way of making sense of 
gravity as a fundamental quantum field theory \cite{reviews}.
One postulates that the Renormalization Group (RG) flow of the theory, 
described by a vector-field in the infinite dimensional space of all possible functionals
of the metric, has a fixed point (FP) with finitely many ultra-violet (UV) attractive directions.
Such directions correspond to so-called ``relevant'' operators and span (the
tangent space to) a finite dimensional surface called the ``UV critical surface''.
The requirement that the theory holds up to arbitrarily high energies then 
implies that the natural world must be described by a RG trajectory lying in this
surface, and originating (in the UV) from the immediate vicinity of the fixed point. 
If the surface has finite dimension $d$, then $d$ measurements performed
at some energy $\lambda$ are enough to pin down all arbitrariness, and then the remaining
(infinitely many) coordinates of the trajectory are a prediction of the theory which can be 
tested against further experiments.

Asymptotically Safe Gravity ({\sc ASG}) is the case in which the fundamental gravitational
interaction is asymptotically safe. This necessitates a modification of General Relativity
whose exact nature is so far unknown. We will not enter here into
a discussion of whether gravity is really asymptotically safe. 
Though we have some evidence \cite{evidence}, 
it is a topic still under discussion. 
For the sake of our argument we will instead just take {\sc ASG} as a given and explore
the connections to {\sc DSR}.

Let $\lambda$ denote the RG scale. From a Wilsonian standpoint, we can refer to $\lambda$
as ``the cutoff''. As is customary in lattice theory, we can take $\lambda$ as
unit of mass and measure everything else in units of $\lambda$.
In particular we define
\beqn
\tilde G=G \lambda^2
\eeqn
the dimensionless number expressing Newton's constant in cutoff units.
(Here and in the following, a tilde indicates a dimensionless quantity.)
The statement that the theory has a fixed point means that $\tilde G$,
and all the other similarly defined dimensionless coupling constants, 
tend to finite values when $\lambda \to\infty$.

For later purposes, it will be useful to have a hint of the general
behavior of the running Newton constant.
Simple dimensional analysis suggests that the beta function of $1/G$ has the form 
\beqn
\lambda\frac{d}{d\lambda}\frac{1}{G}=\alpha \lambda^2~,
\eeqn
where $\alpha$ is some constant. This expectation is borne out by a number of 
independent calculations, showing that the leading term in the beta function
has this behavior, with $\alpha>0$.
Then the beta function of $\tilde G$ has the form
\beqn
\lambda\frac{d\tilde G}{d\lambda}=2\tilde G-\alpha \tilde G^2 ~.
\label{this}
\eeqn
This beta function has an IR attractive fixed point at $\tilde G=0$
and also an UV attractive nontrivial fixed point at $\tilde G_*=1/\alpha$.

Qualitatively the running of Newton's constant is characterized by the
existence of two very different regimes:

\begin{itemize}
\item If $0<\tilde G\ll 1$, which means for subplanckian energies,
the first term on the r.h.s. of Eq. (\ref{this}) dominates. The solution of the flow equation is
\beqn
\tilde G(\lambda)=\tilde G_0\left(\frac{\lambda}{\lambda_0}\right)^2\ ,
\eeqn
where $\lambda_0$ is some reference scale and $\tilde G_0=\tilde G(\lambda_0)$.
Thus, the dimensionless Newton's constant is linear in $\lambda^2$,
which implies that the dimensionful Newton's constant $G(\lambda)=G_0$ is constant.
This is the regime that we are all familiar with.

\item On the other hand in the fixed point regime the dimensionless Newton's constant 
$\tilde G=\tilde G_*$ is constant, 
which implies that the dimensionful Newton's constant runs according to its
canonical dimension: $G(\lambda)=\tilde G_*/\lambda^2$.
%This is the fixed point regime.
In this regime a kind of scale invariance is realized,
in spite of the formal presence of dimensionful couplings.
\end{itemize}

The threshold separating these two regimes is naturally near the Planck scale,
and in the real world $\tilde G$ must 
go from its fixed point value at the Planck scale to very nearly zero
at macroscopic scales.

\section{A Minimal Length?}

Assuming that the behavior outlined above is realized for gravity
(and that all the other couplings go to a fixed point as well)
one can take the limit $\lambda \to\infty$.
Thus, there does not seem to be a maximal energy or a minimal length
scale, in {\sc ASG}. But let us examine this argument more closely.

Just taking the limit of $\lambda \to \infty$ does not tell us there
really is a way to operationally resolving arbitrarily small structures. 
When addressing this question we have to keep in mind that a dimensionful quantity
does not have any intrinsic value: it is only when we measure it
in some unit that we can attribute a value to it.
In other words, if we denote $u$ some quantity having the same dimension
as $\lambda$, and which we take as a unit, then strictly speaking we can only 
measure dimensionless numbers such as $\lambda/u$.

Now, in quantum gravity it is very common and convenient to
measure masses in Planck units\footnote{The argument would not change if we used any other coupling
as a unit, because at a fixed points all dimensionful quantities
run according to their canonical dimension, and any unit of mass
would behave in the same way as the Planck unit.}, so the physically meaningful
``value of $\lambda$'' is the dimensionless number $\lambda\sqrt{G(\lambda)}$.
In the subplanckian regime $G$ can be regarded as a true constant
and this number scales linearly with $\lambda$, as we expect to happen.
But in the fixed point regime, due to the running of $G$,
$\lambda \sqrt{G(\lambda)}$ becomes a constant, namely $\sqrt{\tilde G_*}$.
So even though the dimensionful ``Platonic'' momentum scale $\lambda$
goes to infinity, when measured in Planck units it has an upper bound
\cite{perini3}.
Thus in a very physical sense one can say that the theory has a maximum energy scale
and, by a similar reasoning, a minimum length scale.

It is often said that a minimal length must arise in quantum field theory
when gravity is taken into account, based on very general considerations \cite{Mead:1964zz}.
Perhaps the best known ones are based on some variant of Heisenberg's microscope:
Any attempt at measuring coordinates with precision $\Delta x$
will put in the localization region an energy of order $1/\Delta x$.
When $\Delta x$ is of the order of the Planck scale, $\Delta x$ becomes
comparable to the Schwarzschild radius associated to this energy,
so any attempt at producing a more accurate localization will fail because
a black hole is produced. 
Other arguments examine the attempt 
to measure the location of one particle using a highly energetic test particle. Then, the
gravitational interaction of the test particle itself contributes an additional 
uncertainty to the measurement that becomes relevant only close to the
Planck scale. 
In either case, it is then concluded that one finds an intrinsic fundamental
limit to the possible resolution of structures which  is argued to represent a 
universal cutoff rendering all field theories finite.

Such arguments rely on the validity of Einstein's equations down to
the Planck scale. However, it is not obvious that this is a reasonable assumption
once quantum effects are taken into account.
For example, in {\sc ASG} there is antiscreening:
the dimensionful Newton's constant becomes smaller at shorter length scales,
meaning that the strength of the gravitational coupling does not
increase as strongly as predicted at tree level, 
and eventually completely stops growing.
Recent analysis even indicates that, under these conditions, Planck scale black holes
do not form \cite{falls}.

The argument we have offered instead is of purely quantum nature, but interestingly 
still leads to the same conclusion without assuming the validity of the
classical Einstein's equations in the Planckian regime. This relation will
be further studied in a forthcoming publication \cite{vacca}.

\section{DSR from Asymptotically Safe Gravity}

In the preceding section we have argued that the physically meaningful, dimensionless,
quantity $\tilde {\bf p}$ obtained from any momentum or energy ${\bf p}$ 
is the one being divided by the energy-dependent
Planck mass: $\tilde {\bf p} = {\bf p} \sqrt{G(\lambda)}$,
where $\lambda$ is some as yet unspecified function of ${\bf p}$. 
For the sake of interpretation it is however more 
useful to reconnect this dimensionless quantity with another
momentum %${\bf k}$, 
measured in constant units. 
One obtains the ``physical'' momentum ${\bf k}$ from the ``Platonic'' 
momentum ${\bf p}$ by rescaling with
the ratio of the fixed over the running Planck mass:
\beqn
{\bf k} = {\bf p} \sqrt{ \frac{G({\lambda})}{G_0}} ~. \label{kp}
\eeqn
As we argued in the preceding section,
the assumption of asymptotic safety then implies that this
physical momentum ${\bf k}$ has an upper bound. 
If we take $\lambda$ to be some function of the momentum itself, 
for example $\lambda=p^0$, the similarity to
{\sc DSR} is striking, though we have to be careful not to jump to conclusions. 
The interpretation of the above equation depends on
how one identifies the energy scale $\lambda$ in the particular process one is
considering.

It is important to note that the factor $G({\lambda})/G_0$ in Eq. (\ref{kp}) 
does not have to be invariant under the standard Lorentz transformation, 
i.e. is not necessarily an ordinary Lorentz scalar. 
It might for example contain contractions of momenta with the metric or curvature 
components, which are also functions of the energy content of the fields. 
If one makes a Taylor-series expansion of Eq. (\ref{kp}) in ${\bf p}$, the
coefficients have a transformation behavior that is mixed between standard
special relativity and {\sc DSR}.
In principle, one could use Eq. (\ref{kp}) for each specific process and
derive the form of {\sc DSR} one is dealing with. However, our current understanding 
of {\sc ASG} is not sufficiently developed to do this. In \cite{girelli} it was studied 
how a modified dispersion relation can arise from the renormalization group of gravity.
It was assumed there however that the modification is a function of the
three momentum squared. 
 
Here, we will instead examine the interpretation of the {\sc DSR} we arrive at. 
For that, we will consider two thought experiments:
the measurement of the momentum of a free particle, 
and then a highly energetic scattering process. 

In the {\sc ASG} interpretation of {\sc DSR}, for a freely propagating particle, there can be 
no {\sc DSR}-modification, since there is nothing setting a scale that is being probed. A
free particle does not cause strong gravitational effects, thus the coupling constant
is well within the classical regime and the conversion factor in Eq. (\ref{kp}) is equal
to one to excellent precision. 
But in \cite{Liberati:2004ju} it was suggested to understand {\sc DSR} as an effect 
occurring in the measurement process, an interpretation that we wish to investigate
here. The momentum ${\bf p}$, transforming linearly 
under Lorentz transformations, is then an intrinsic property of a particle,
while ${\bf k}$ is the result of the measurement and
the function $F$ that relates the two is a property of the apparatus,
taking into account the properties of the gravitational interaction between the 
particle and the apparatus. 

To connect this thought to our approach from {\sc ASG}, the question we are facing here 
is what would we measure for the momentum of a free particle. The notion itself is an 
oxymoron of course. A measurement is an interaction process, and thus the particle
is no longer free when measured. To come as close as possible to the desired goal, we 
consider a measurement process taking place through many soft interactions. The concrete
example we will have in mind is that of a  particle detector in which
the track of the particle,
from which its momentum is deduced, consists of
a very large number of ionization events, each removing a very
small amount of momentum from the particle. But that then means that
the relevant value of $\lambda$ would always be small in Planck units
and therefore this process would not probe the Planckian regime at all, 
even if the momentum of the particle (relative to the
detector) is in the Planckian regime. Another way to see that {\sc DSR} cannot arise in this way from the measurement process 
is that the measurement
does not necessarily have to take place via gravitational interaction
and certainly not via hard graviton exchange.

Consider now a scattering process, $s$-channel at tree-level, in the Planckian regime. The
initial and final states are standard model particles, i.e.\ there is no graviton emission. 
The question is, what is the effect we get from the gravitational interaction that 
becomes non-negligible in this energy regime?

As argued above, for the ingoing and outgoing asymptotic scattering
states we have no {\sc DSR}-modification and also no bound on the momentum. Things do 
change however in the collision region. The scattering particles create a virtual
particle in a region of strong gravitational effects. The scale that we have available
here is the momentum of the exchange-particle, or the total momentum of the initial
particles. It is this scale at which the running coupling should now be
evaluated, leading us to conclude that the physically relevant momentum of the
exchange particle is lowered in the {\sc ASG} scenario from ${\bf p}$ to ${\bf k}$ with
$G$ being a function of the total energy of the initial particles. The virtual 
particle then produces asymptotically outgoing scattering states for which again we have
no modification. 

We can interpret this as follows. When the initial particles approach each
other, quantum gravitational corrections become non-negligible. In this
region, the gravitational interaction is modified in such a way
that energy is effectively redistributed into gravitational degrees of freedom, 
leading to a decrease in the momentum that is transferred in the actual particle
collision. If one merely looks at the interacting particles, it seems as
if energy is decreasing when quantum gravitational effects become relevant,
and is increasing again when they cease to be relevant. Since we have excluded 
real emission in gravitons however, the total energy of the scattering process 
is conserved, meaning that (by assumption of the scenario considered) the
redistribution of energy is only temporary and confined to the interaction
region. (Since for the in- and outgoing particles there is no
modification, there is no ambiguity in identifying the conserved quantity.)

This interpretation is supported by the analysis in \cite{Bonanno:2010bt},
where it was found that in the RG-improved field equations of gravity,
energy-momentum may not be conserved a prima facie. If one computes
the Einstein tensor, one of course obtains an effective energy-momentum tensor 
that is conserved as usual. But then this effective tensor contains
contributions from the background geometry. These contributions can be
interpreted as a fluid which interacts with the coarse grained gravitational degrees of 
freedom. 

The interpretation of {\sc DSR} that we arrive at here from the {\sc ASG}-approach 
differs from the usual {\sc DSR} scenario in that we have no modification for freely propagating particles,
basically because there is nothing besides the invariant mass of the particle 
that could set a scale for this modification, and for all particles we know these
masses are much below the Planck scale. The interpretation proposed here 
differs from the interpretation in \cite{Liberati:2004ju} in that we have found
the measurement process to not cause {\sc DSR}-modifications either.
The interpretation found here is instead akin to the one considered in
\cite{Hossenfelder:2003jz,Hossenfelder:2006cw} in that here too a modification is only
present for virtual particles. These two scenarios differ however in
two points. First, in \cite{Hossenfelder:2003jz,Hossenfelder:2006cw}, the quantity ${\bf k}$
has been interpreted as the wave-vector of the virtual particle,
an interpretation for which we have no need here. Second, the model
considered in \cite{Hossenfelder:2003jz,Hossenfelder:2006cw} is based on
one universal function relating the quantities ${\bf k}$ and ${\bf p}$. Here instead,
the relation between these both quantities is dependent on the process under
investigation.  

The interpretation of {\sc DSR} suggested by the {\sc ASG} approach has
the merit of doing away with the two major problems of the standard
{\sc DSR}-models -- the construction of multi-particle states and
nonlocal effect -- for the same reason that the model considered in
\cite{Hossenfelder:2003jz,Hossenfelder:2006cw} does not suffer from
these problems. There is, in the interpretation suggested here, no
problem with macroscopic nonlocality because there is no modification 
for the propagation of free particles. The model still has nonlocal
effects at the Planck scale, but
these cannot be amplified by travel over long distances since modifications
are only relevant in interaction regions. There is no problem with obtaining
multi-particle bound states because here the energy relevant for the
strength of the {\sc DSR}-modifications is not given by the total energy
of the state, which can easily exceed the Planck mass, but by the typical
binding energy, which is far below the Planck scale.

\section{Conclusion}

We have found that asymptotically safe gravity, while not having
an explicit notion of a minimal length scale, does have an implicit
notion of a minimal length that can become relevant for
physically measureable quantities. We have investigated the implications
of this by considering two thought experiments and concluded that there
is no modification for free particles, but scattering processes in
the superplanckian regime will be modified. The key difference of
the approach discussed here to {\sc DSR} and other approaches is that the modification one 
obtains here is dependent on the process one considers. The 
approach to {\sc DSR} from asymptotically safe gravity presented here is promising as it brings with it 
the prospect of obtaining a concrete realization of {\sc DSR}.

\medskip
\section*{Acknowledgements}

SH and RP thank the Perimeter Institute for hospitality
during the early stages of this work. This work is supported in part by the European Cooperation in Science and Technology ({\sc COST}) action {\sc MP0905} ``Black Holes in a Violent Universe.'' 

\noindent

\goodbreak

\end{document}